\documentclass[
reprint,
superscriptaddress,
amsmath,amssymb,
prstab,
]{revtex4-2}

\usepackage{placeins}
\usepackage{enumerate}
\usepackage{csquotes}
\usepackage{graphicx}% Include figure files
\usepackage{amsmath}
\usepackage{subcaption}

\usepackage{siunitx}
\DeclareSIUnit\bar{bar}
\DeclareSIUnit\var{VAR}

\usepackage{tablefootnote}
\graphicspath{ {./images/} }
\usepackage{xargs}                      % Use more than one optional parameter in a new commands
\usepackage[pdftex,dvipsnames]{xcolor}  % Coloured text etc.

\begin{document}
\preprint{AIP/123-QED}
\title{Detailed Design and Optimization of Ferro-Electric Tuners}
\author{Ilan \surname{Ben-Zvi}}
\email{Ilan.Ben-Zvi@StonyBrook.edu}
\affiliation{Physics and Astronomy Department, Stony Brook University, NY USA}
\author{Alick \surname{Macpherson}}
\email{alick.macpherson@cern.ch}
\affiliation{CERN, CH-1211 Geneva, Switzerland}
\author{Samuel \surname{Smith}}
\email{Samuel.jack.smith@cern.ch}
\affiliation{CERN, CH-1211 Geneva, Switzerland}

\date{\today}

\begin{abstract}
A detailed, step-by-step design methodology of a Ferroelectric Fast Reactive Tuner (FE-FRT) capable of modulating Mega VAR reactive powers on a sub-microsecond time scale is given. Closed expressions of values for all the components of the tuner are detailed, and tuner performance optimization is addressed, resulting in a Figure of Merit measure of FE-FRT tuner performance and use case applicability over a wide range of RF frequencies and reactive power levels. This enables addressing feasibility  and rapid assessment of  design parameters given an FE-FRT tuning scenario defined by required tuning range, cavity operating frequency a cavity stored energy. 
\end{abstract}

\maketitle

%\pagebreak

\section{Introduction}\label{section1}

Frequency tuning of accelerator cavities is a fundamental tool of accelerator science and technology. The Ferroelectric Fast Reactive Tuner (FE-FRT) extends the capability of tuners towards faster and higher-power tuning  for both normal conducting and superconducting cavities. This communication extends the previously developed \cite{ ben2024conceptual,BenZvi2021HighpowerFF} theory and methodology of high-power FE-FRTs and adds a detailed, step-by-step design procedure for such an FE-FRT device and provides methods for optimizing its performance over a  broad range of RF frequencies and reactive power levels. Finally, the reader is  guided through the electromagnetic simulation procedure, so to permit comparison between the analytic model and electromagnetic design. 

In addition to the  presentation of the FE-FRT design process,  one of the main conclusions of this work is the design advantage gain from  FE-FRT tuner designs comprising of just two ferroelectric wafer elements. Given the challenge of designing high average power tuners, where power dissipation in the ferroelectric material is a central design driver, the  interplay between power handling capability  and the RF design is what is developed in this paper, in terms of FE-FRT applicability and tuner performance reach.  The rationale for key design choices to determining  optimum tuner parameters compliant with given frequency and tuning power requirements is given, and is followed by a comparison of analytical results with detailed numerical simulation of several challenging high power tuner scenarios. The comparisons show that analytic expressions, evaluated using Maple \cite{maple2019} closely approximate results obtained with CST’s Studio suites' finite element solver \cite{CST}, giving credence to the underlying design methodology.

\subsection{Methodology}\label{Methodology}
When tasked with designing a high power tuner, the typical inputs constraints are: 
\begin{itemize}
    \item The cavity resonant frequency, $f_0$, $\omega_0 \equiv 2\pi f_0$.
	\item The required tuning range, $\Delta f$.
	\item The stored energy of the cavity, $U$.
\end{itemize}
The tuner can be considered as an impedance at the port of the cavity to be tuned, with the complex power $P$ flowing from the cavity to the tuner. With this view, the real part of this complex power is dissipated power in the tuner, which has to be accommodated for in its design. Naturally, this dissipated power should be kept low so thermal management of the tuner is less demanding, but also to reduce undesirable draining of RF power from the cavity that would have to be replenished by the RF drive. The imaginary part of $P$ is the reactive power flowing into the tuner, and is responsible for the frequency tuning of the cavity. For a tuning range (lowest and highest frequency) defined by "tuner end states" and described with subscripts 1 and 2, the product of the stored energy and the required tuning range determines the reactive power that has to flow from the cavity to the tuner \cite{ ben2024conceptual}:
\begin{equation}
    \Im (\Delta P_{12} ) = 2U \Delta \omega_{12} 
    \label{eq1}
\end{equation}

 For demanding tuner design requirements, a large reactive power is to be foreseen,  and  this in turn can imply additional engineering considerations to the design, in order to avoid issues such as RF voltage breakdown thresholds at the coupler port, transmission line or tuner body. 

In order to assess the performance of such a reactive tuner, a Figure of Merit ($FoM$) can be defined to quantify the ability of the tuner to produce reactive power for tuning a cavity given limits on the either power dissipation in the tuner circuit or power drawn from the cavity. The $FoM$ represents the maximal change in reactive power per the average dissipated power in the tuner system, and  is a clear performance indicator for the tuner. The $FoM$ can be presented in terms of the reactance $X$ (used for the frequency tuning) to the resistance $R$ (wasted dissipation) over the full tuning range, where $Z= R + jX$ is the complex impedance of the tuner at the cavity port. The $FoM$ is given by: 
\begin{equation}
    FoM \equiv \frac{\Im(P_2 )-\Im(P_1 )}{2\sqrt{\Re(P_1 )\Re(P_2) }} 
        = \frac{X_2-X_1}{2\sqrt{R_1 R_2}}
    \label{eq2}
\end{equation}
Given that the reactance and resistance vary in a complex way as a function of the permittivity, the $FoM$ as defined is a useful but approximate indicator of performance. 

\section{Key design choices}\label{section2}
Until now, the details of the tuner circuit were not determined, and the tuner is simply represented by a tunable impedance at the port of the cavity. Assuming a tunable reactance achieved by a ferroelectric loaded capacitor for  tuning functionality, reference \cite{ ben2024conceptual} introduced a number of potential equivalent circuit designs, with  both non-resonant  and resonant circuits options considered. 

\subsection{The Tuner as an Impedance} \label{section2_1}

\begin{figure}[tb]
    \centering
    \includegraphics[width=0.95\columnwidth]{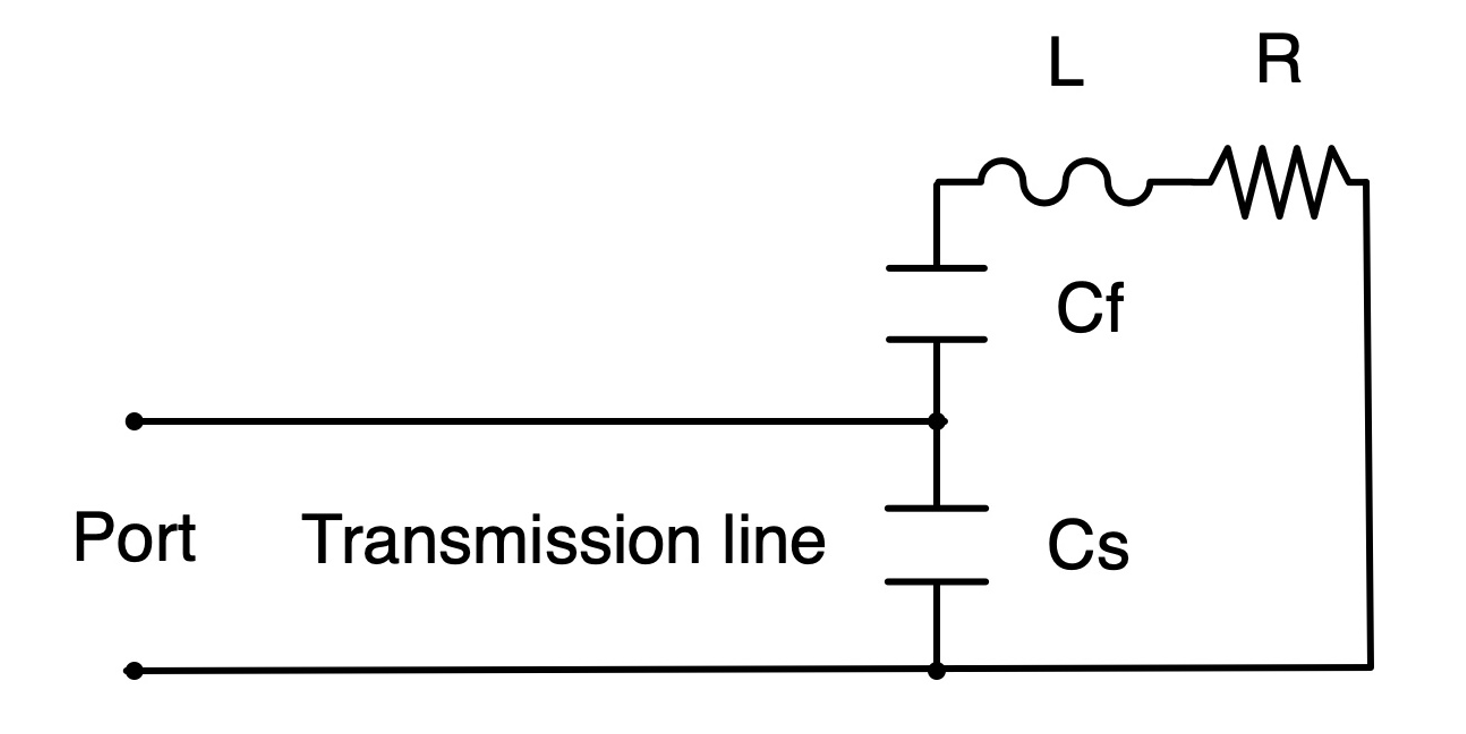}
    \caption{\label{Circuit} The equivalent circuit of a capacitive coupled resonant tuner.} 
\end{figure}

\begin{figure}[tb]
    \centering
    \includegraphics[width=0.95\columnwidth]{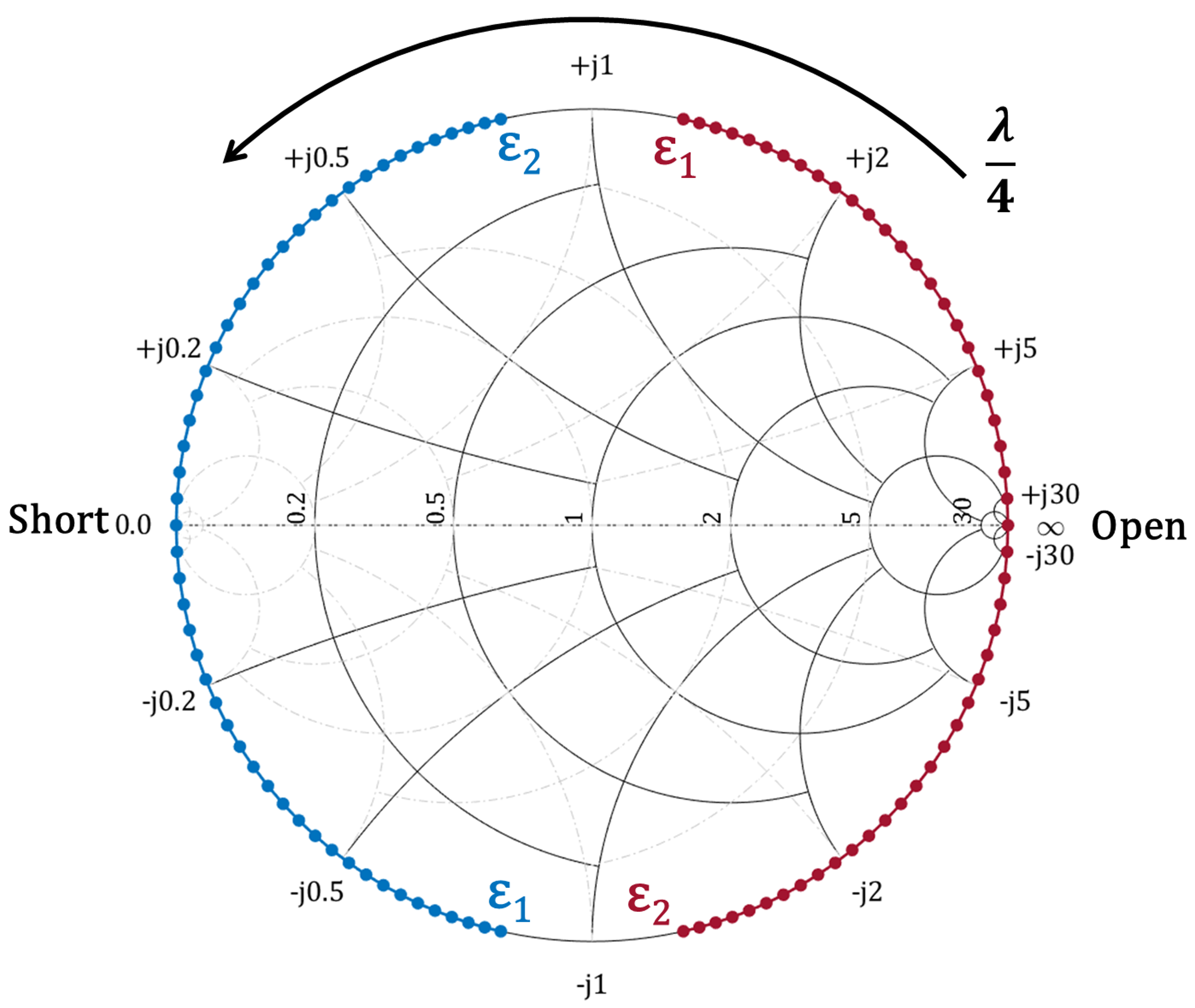}
    \caption{\label{smith} Smith chart showing an example of the tuner's impedance as a function of permittivity centered about a central permittivity for just the resonator (marked in red) and for the resonator followed by a quarter-wave length transmission line (marked in blue).} 
\end{figure}

For achieving a high $FoM$ performance, a resonant tuner circuit with capacitive coupling to the cavity was found to be the most effective, and is taken as the starting point for this work. Fig. \ref{Circuit} gives the general equivalent circuit, where the $C_f$, $L$, $R$ are the resonator parameters containing the ferroelectric segment, with $C_S$ is for impedance matching  of tuner to the cavity. While the resonant frequency of the tuner circuit is set by the LC circuit, the choice of tuner inductance $L$ can be made such that the tuning range is made symmetric around $f_0$,  thereby minimizing the peak voltage. Also implicit in this schematic is a transmission line, with line length set at one quarter wavelength, which transforms the tuner impedance such that at the interface to the cavity, the tuner system presents an impedance that is well defined and continuous over the full tuning range,  thus  symmetric around the short position on the Smith chart (see Fig. \ref{smith}). This configuration is appropriate for an analog tuning response.


As previously established \cite{ben2024conceptual}, the achievable tuning range can be derived by considering the tuner as an external impedance coupled to the cavity with transformer turn ratio given by a port impedance $Z_0$ and  a  quality factor $Q_e$:
\begin{equation}
        \Delta \omega_{12} = 2 \pi \Delta f_{12} = \frac{\omega_{0}}{2 Q_e} \frac{X_2-X_1}{Z_0} 
    \label{eq3}
\end{equation}
For a given tuning range, the ratio of the change in reactance $X_2-X_1$ to the product $Q_e Z_0$ is determined, and in what follows, the procedure for determining the required change in reactance, $X_2-X_1$, that maximizes the $FoM$ is detailed.

\subsection{The Ferroelectric Material}\label{section2_2}

In the FE-FRT tuner the tuning function is  achieved through the application of a variable bias voltage to a ferroelectric capacitor, with the applied bias voltage inducing a change in the permittivity of the ferroelectric. This tunable permittivity implies a controllable variable reactance that can be incorporated into the resonant LC circuit of the tuner, and coupled to the cavity. To realize this ferroelectric capacitance, a flat wafer format is ideal for the ferroelectric material, as it permits a reduced bias voltage for a given polarizing electric field (small wafer thickness) and improves the thermal conductivity for cooling the wafer (larger wafer surface area). In addition, reference \cite{ ben2024conceptual} demonstrated that an annular wafer geometry reduced parasitic inductance and resistance losses at the capacitor-spacer transition. As such, the ferroelectric capacitance considered here is a stack of $N_w$ of annular wafers  connected in series, with the wafers  separated by copper spacers  that allow both for the application of bias voltage across the ferroelectric wafers and for the provision of cooling of the wafers.

Assuming  an annulus wafer geometry in a multi-stack assembly of ferroelectric capacitors, the total capacitance area  of the wafer stack of ferroelectric capacitors is $A=N_w \pi (r_a^2-r_i^2)$ and is the surface area over which heat can be extracted from the ferroelectric. Here, $r_a$ and $r_i$ the outer and inner radii of the wafers, and the wafer thickness is $g$. 

The tuning range is  mapped to a range in relative permittivity $\epsilon$ of the ferroelectric, where $\epsilon$  is changed by applying a bias electric field across the ferroelectric capacitor, with the full tuning range is mapped to a relative permittivity range $\epsilon_1 - \epsilon_2$  defined by the tuner end states.  

In terms of the current ferroelectric of choice, a BST ceramic ((BaTiO3/SrTiO3-Mg) \cite{kanareykin2005low} is recommended, which  has a room temperature of permittivity  of $\sim\epsilon=160$, and is a low loss ceramic, with a very low frequency dependent loss tangent. The typical measured loss tangent value is $\delta = 3.4 \times 10^{-4}$ at \SI{80}{\mega \hertz}, and a frequency scaling as $\sim \omega ^{0.783}$ between \SI{80}{\mega \hertz} and \SI{400}{\mega \hertz}, depending on material  temperature  has been reported \cite{kanareykin2005low}. 

Given that power dissipation in the ferroelectric material dominates the technical challenges when designing high-power tuners a  $FoM$ for bare ferroelectric material can be derived,  and as it neglects any dissipation in other elements of the tuner circuit, and this represents an upper limit on the achievable $FoM$. This reference $FoM$ is given by  
\begin{equation}
    FoM(FE)=\frac{\Delta \epsilon}{2 \delta \epsilon_c}
    \label{eq4}
\end{equation}
where $\Delta  \epsilon=|\epsilon_2 -\epsilon_1|$, and $\epsilon_c$ is a central value for the relative permittivity, approximated by $\epsilon_c \approx \sqrt{\epsilon_1 \epsilon_2 }$. For the aforementioned BST ceramic,  this figure of merit is temperature dependent with a  peak at about \SI{50}{\degreeCelsius}. At this temperature, with permittivity end-states defined by a sustainable bias electric field of up to 8 MV/m,  $\epsilon_1=96, \epsilon_2=130$, and the frequency dependent loss tangent at \SI{400}{\mega \hertz} is  $\delta =9.5 \times 10^{-4}$. This in turn implies a $FoM(FE)=156$, which can be considered an upper limit on tuner performance. For  higher power tuner designs, Eq. (\ref{eq4}) implies that a lower average permittivity enables an improved $FoM(FE)$ performance reach, but may  come at the cost of limited the permittivity  range, and thereby limiting tuning range.

\subsection{Aspect Ratio of the Ferroelectric Capacitor}
To understand the implications of power dissipation to the realization of the tuner design geometry, the power dissipated in the tuner’s circuit is distributed between the ferroelectric ceramic and the copper conductor elements and has to be considered separately. For a ferroelectric capacitor  comprised of $N_w$ wafers connected in series, the wafer geometry is key to optimizing the power dissipation, with the wafer area-to-thickness aspect ratio used to quantify both the ferroelectric capacitance and the  thermal transport at the  wafer interface. The later is relevant, as it determines the temperature rise in the wafer that must be accommodated for by  a cooling system integrated into the wafer stack spacers.  For example, assuming the cooling circuit provides a refrigerant at \SI{20}{\degreeCelsius}, and the ferroelectrics operating temperature is set at  \SI{50}{\degreeCelsius}, the maximum allowable average temperature increase of the ferroelectric from dissipative losses  is limited  to \SI{30}{\degreeCelsius}. This limitation is not strictly necessary but represents a compromise between maximizing the $\text{FoM}(\text{FE})$, minimizing the wafer's area, and addressing the engineering constraints of the cooling circuit. This is particularly relevant to the cooling circuit geometry, as for a fluid cooling system that depends on the ability to extract dissipative power from the spacers between the wafers, mechanical constraints from cooling channel dimensions must be considered; as a conservative estimate, the spacer thickness must be at least 5 mm for $N_w=2$ or 10 mm for larger $N_w$ to accommodate cooling lines.  It is worth noting that other methods for cooling the wafers could be envisioned, but the analytic equations provided herein  are independent of the specifics of the implementation details.

\begin{figure}[tb]
    \centering
    \includegraphics[width=1\columnwidth]{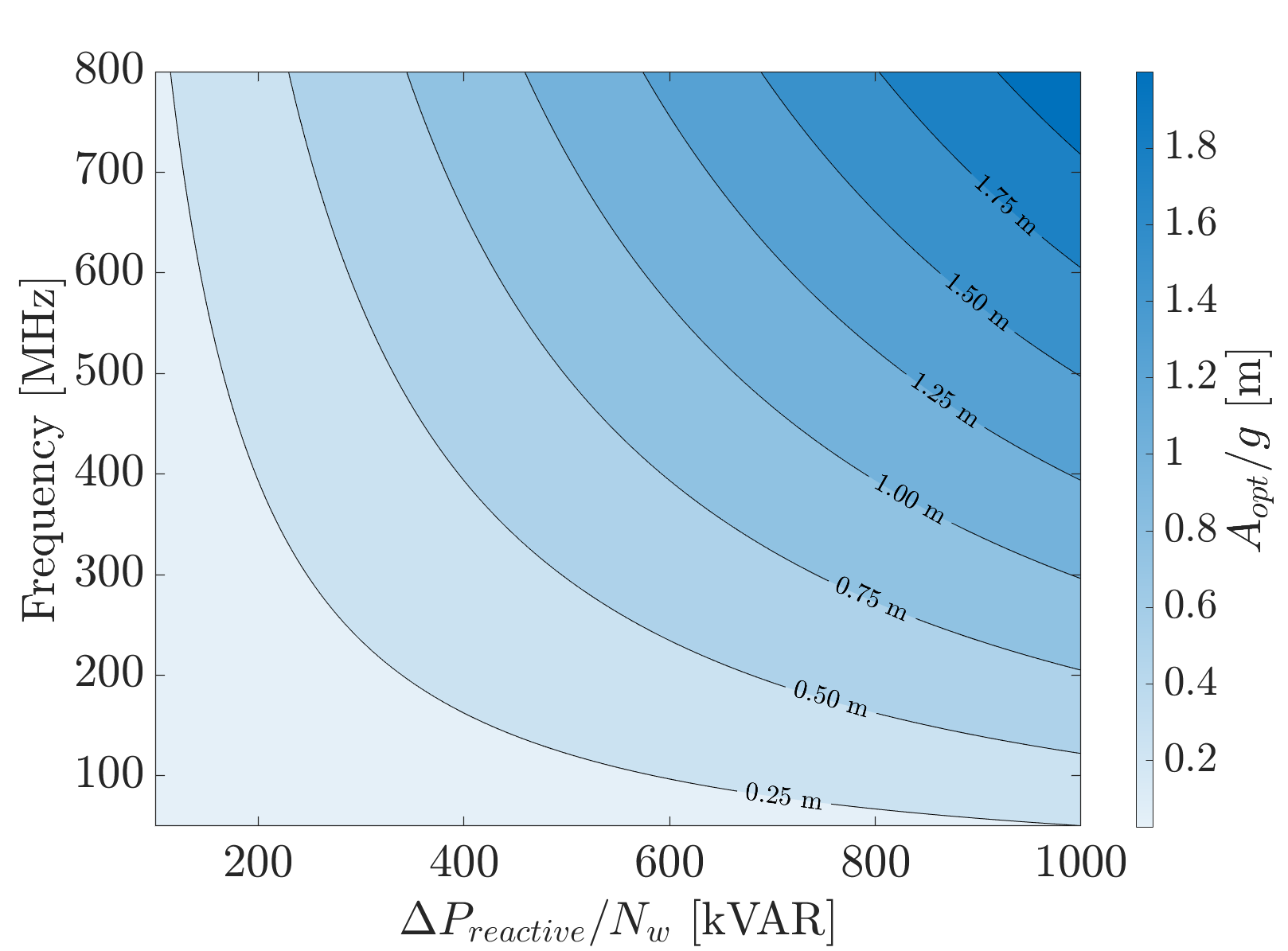}
    \caption{\label{A_over_g} The optimal ratio of a single wafer cross sectional area to gap $A_{opt}/g$ in meters  as a function of   the wafer reactive power $\Delta P_R/N_w$ in units of kVAR and operating frequency in MHz. The ferroelectric material properties are as noted in section \ref{section2_2}, and a wafer temperature rise of 30 degrees is assumed.}
\end{figure}

To determine the optimum aspect ratio of an individual wafer, with a single-face surface area of $A$, consider that tuner operation generates a power dissipation of $P_W$ in the $N_w$ wafer series-connected stack. Assuming that heat can be extracted from both faces of each wafer, and that the cooling can accommodate an average temperature rise $\Delta T$ in the wafers, then for a power flow of $P_W/2$ flowing from the wafer center to the surface area $A$
\begin{equation}
    \begin{aligned}[c]
     %   P(y) &= \frac{P_W}{2}(1 - \frac{2y}{g})=KA\frac{dT(y)}{dy} \\
        \Delta T&=\frac{2}{g} \int_0^{g/2} T(y) dy = \frac{P_W}{12KA}
    \end{aligned}
    \label{T_Aopt}    
\end{equation}
Here  $K$ is the thermal conductivity  of the ferroelectric material \cite{kanareykin2005low}. Equating this $\Delta T$ with that of the target temperature rise in the wafer relative to its cooled surface,  the optimal wafer aspect ratio $A_{opt}/g$ can be given by:
\begin{equation}
    A_{opt}/g=\frac{P_W}{12 \Delta T K} 
    = \frac{\Delta P_R}{12 N_w K \Delta T}   \frac{ \delta (\omega)\sqrt{\epsilon_1 \epsilon_2 }}{\Delta \epsilon}
    \label{eq5}
\end{equation}

where $\Delta P_R=\Im (\Delta P_{12}) =2N_w P_W FoM(FE)$  is the peak reactive power in the complete tuner.  Given that the FoM improves with a reduced capacitance, this aspect ratio is  considered as optimal, as the wafer stack capacitance is proportional to $A/(gN_W)$.  Fig. \ref{A_over_g} shows a contour plot of this optimal aspect ratio as a function of  peak reactive power in a single wafer $\Delta P_R/N_w$ and cavity frequency $f_0$. 

With an optimal ratio $A_{opt}/g$, the actual area and wafer thickness must be determined by engineering considerations,  which are beyond the scope of this paper. However, it is noted that while the power of the bias driver of the ferroelectric capacitor varies as $g^2$ giving a preference for a smaller $g$, smaller wafer areas may increase the temperature step from the spacers to the wafers, making cooling them more difficult, and  suggesting a larger value of $g$ is preferred. As a compromise, a  wafer thickness between 0.2 mm and 1 mm is recommended, depending on manufacturing feasibility.

\section{THE RESONANT TUNER}\label{section3}
For  optimum tuner performance, the tuner equivalent circuit shown in Fig. \ref{Circuit} should be resonant over frequencies defined by the cavity and required tuning range, and   provides two important advantages over non-resonant designs: \begin{itemize}
    \item  The reactance change of the ferroelectric capacitor is amplified parametrically to yield a reactance change of the resonator which is $Q$ times larger, where $Q$ is the quality factor of the resonator. 
    \item By tuning the resonator's frequency $\Omega_0$, taken at a central value of the permittivity range to about the cavity's frequency $\omega_0$, the impedance of the resonator is made symmetric in both end states of the tuner.
\end{itemize}

 In the following sections, the constraints on the $LCR$ circuit parameters are detailed.
 
\subsection{The Modeling the Resonator}

With the $LCR$ circuit parameters as per Fig. \ref{Circuit}, $C_f$  represents the ferroelectric capacitance, $L$ the tuner inductance including the ferroelectric capacitor and its support spacers, and  $R=R_C+R_L$, is the the total resistance of the circuit, comprised of a contribution from dielectric loss in the capacitor as well as the inductive resistance. The resonant frequency of the tuner's circuit is then $\Omega = 1/\sqrt{LC}$, where $C$ is the total capacitance, and the resonant circuit has a shunt resistance $\mathbb{R}=Q \sqrt{L/C} $. The total tuner capacitance  $C$ is the in-parallel sum
 \begin{equation}
    C = \frac{C_s C_f}{C_s + C_f}
    \label{eq26}
\end{equation}

and given the dissipative nature of the ferroelectric capacitor, it contributes $R_C=\delta/(\Omega_0 C_f )$ to the resistivity. 

To quantify the tuner inductance $L$ for a stacked wafer geometry, the inductor may be considered as a short coaxial transmission line. At low frequencies ($f_0 <$ \SI{30}{\mega\hertz}) the resonator can be constructed as a compact lumped element resonant circuit, but this simplification is not relevant to most tuners, so the short terminated coaxial line modeling is adopted. 

To fully model the total resistance of the resonant section, the inductor loss is taken as the series sum of the resistive loss from the ferroelectric, the resistance of the short transmission line segment, and the effect of the current at the end-wall terminations of the line representation of the inductor, so that 
\begin{equation}
    R = R_C +R_L = R_C+ R_i+ 2 R_B
    \label{eq9a}
\end{equation}

where $R_B$ is the end-wall line termination resistance of the resonator. Note that the inductor is modeled as a coaxial line terminated at both ends, so  $R_B$ appears twice in Eq. (\ref{eq9a}).

Using the standard transmission line expression for the impedance of the  coaxial  stacked wafer $LCR$ resonator the resistance contribution $R_i$ can be determined  as
\begin{equation}
    \begin{aligned}[c]
        Z_{i} = & Z_{r0}  \frac{2R_B+Z_{r0} \tanh (\gamma l)}
                      {Z_{r0}+2R_B \tanh(\gamma l_r)} \\
        & \sim 2R_B +Z_{r0} \alpha l_r(1+\tan(\beta l_r)^2) +j Z_{r0}   \beta l_r  \\
        & = R_i +2R_B +j X_i
    \end{aligned}
    \label{eq7}
\end{equation}
where $Z_{r0}$ is the characteristic line impedance for the resonator,  $\gamma = \alpha + j \beta$ is the complex propagation coefficient of the line, with $\beta =\Omega/c$, and  $l=l_r$ is the transmission line length. From this, the resistive component $R_i$ is:

\begin{equation}
    R_{i} \sim Z_{r0} \alpha l_r (1+\tan(\beta l_r)^2)
    \label{eq10}
\end{equation}
 where the approximation assumes the tuner is compact, and $l_r$ is much shorter than the wavelength of the RF.

Here the transmission line attenuation factor $\alpha$ is defined for a coaxial geometry of with inner and outer conductor radii of $a_r$ and $b_r$ respectively,  as 
\begin{equation}
    \alpha =  \frac{R_s}{2\eta \log(\frac{b_r}{a_r})}(\frac{1}{a_r}+\frac{1}{b_r})
    \label{eq9}
\end{equation}
where $\eta =\sqrt{\mu_0/\epsilon_0 } =376.7 \Omega$ is the impedance of free space,  $R_S=\sqrt{\pi f\mu_0/\sigma}$ is the the surface resistivity, and $\sigma$ is the bulk conductivity of the conductors.

Similarly, the effect of the current at the end-wall of the transmission line, represented by $R_B$, for a standard coaxial short is given by
\begin{equation}
    R_B=\frac{R_S}{2\pi } \log(b_r/a_r)
    \label{eq9b}
\end{equation}

This now permits evaluation of the impedance $Z_r$ of the tuner section, as a capacitively coupled  resonator as measured at the tap point of the transmission line, and is a series-parallel combination, with 
\begin{equation}
        \begin{aligned}[c]
    Z_r & = \frac{Z_{C_S} (Z_{C_f}+Z_i +R)}{Z_{C_S} +(Z_{C_f}+Z_i +R)} \\
        & =\frac{\frac{1}{j\Omega C_s }  (
        \frac{1}{j\Omega C_f } + \frac{\delta}{\Omega C_f }  +R_i+j \Omega L)}{\frac{1}{j\Omega C_s } +
        \frac{1}{j\Omega C_f } + \frac{\delta}{\Omega C_f }  +R_i+j \Omega L}  \\
        & = -\frac{1}{1+ j \frac{2Q\Delta\Omega}{\Omega}} \frac{Q}{\Omega C_S}[\frac{C_f}{C_s+C_f} +\frac{j}{Q}] \\
        & \approx -\frac{\mathbb{R}}{1+ j \frac{2Q\Delta\Omega}{\Omega_0}} (\frac{C_f}{C_s+C_f})^2
    \label{eq13}
    \end{aligned}
\end{equation}
Here  the shunt resistance of the resonator is $\mathbb{R} = \Omega L Q$ and $\Delta \Omega= \Omega- \Omega_0$  is the difference in tuner resonant angular frequency compared to its average permittivity  value $\Omega_0$ which is a function of the permittivity change ($\Delta\epsilon= \epsilon-\epsilon_0$) around the permittivity associated with $\Omega_0$. Balanced operational performance of the tuner implies the center frequency of the tuner's resonator $\Omega_0$ is set to $\omega_0$. The approximation in Eq. (\ref{eq13}) is due to the strength of the quality factor $Q$, and shows that matching the impedance of the resonator toward the cavity is controlled by the ratio $C_f/(C_s+C_f)$.

Neglecting losses in the quarter-wavelength transmission line connecting the tuner to the cavity, the resonator impedance $Z_r$ is transformed by the standard transmission line equation to give the tuner load impedance  $Z_L$ at the connection to the cavity port:
\begin{equation}
    \begin{aligned}[c]
    Z_L &=R_L(\epsilon)+jX_L(\epsilon) \approx \frac{Z_0^2}{Z_r} \\ 
        & = Z_0^2  \frac{1+ j Q \frac{C_s}{C_s+C_f} \frac{\Delta\epsilon}{\epsilon}}
        { \mathbb{R}{(\frac{C_f}{C_s+C_f} )^2}}\\
        & R_L =\frac{Z_0^2}{\mathbb{R}}(\frac{C_s+C_f}{C_f})^2\\
        & X_L = Z_0^2 \Omega_0 \frac{C_s^2}{C_f}  \frac{\Delta \epsilon}{\epsilon} 
    \label{eq17}
    \end{aligned}
\end{equation}
with $\mathbb{R}/Q=1/\Omega_0 C$, and where $Z_0$ is the characteristic impedance of the quarter-wavelength transmission line. Here the finite derivative  of $\Omega^2= 1/LC$ with respect to $C$ has been used to give the detuning relation 
\begin{equation}
    \frac{\Delta\Omega}{\Omega} = -\frac{C_s}{C_s+C_f} \frac{\Delta\epsilon}{2\epsilon_c}
    \label{eq16}
\end{equation}

Given the complex tuner impedance $Z_L=R_L + j X_L$  from above, and the definition of the FoM in Eq. (\ref{eq2}),  by assuming a symmetric tuning $|\Delta \Omega_{10} |\simeq |\Delta\Omega_{20} |$ and noting that at the extremal points  $(2 Q\Delta\Omega/\Omega_0)^2\gg 1$, the  tuner $FoM$  at the cavity  can be expressed as \begin{equation}
    FoM= \frac{\Delta X_{12}}{2\sqrt{R_1R_2}} \approx\frac{C_s}{(C_s+C_f)} Q \frac{\Delta \epsilon_{12}}{2\epsilon_c}
    =\frac{C}{C_f} Q \frac{\Delta \epsilon_{12}}{2\epsilon_c}
    \label{eq20}
\end{equation}
and shows that the $FoM$ depends on the relative strength of the ferroelectric capacitance $C_f$ to the series capacitance $C_s$ and  to the quality factor $Q$ of the resonant tuner, but is independent of the tuner shunt impedance $\mathbb{R}$.  

The $FoM$ can be further  simplified, as the quality factor $Q$ can be expressed in terms of circuit quantities; taking the the resonator resistance as parameterized in terms of the ferroelectric material and coaxial modeling of the conductor losses, the  quality factor of the tuner $Q$ of the resonator segment of the tuner is 
\begin{equation}
    \begin{aligned}[t]
	    Q & =  \frac{1}{\Omega_0 C R} = \frac{1}{ \Omega_0 C(\frac{\delta}{\Omega_0 C_f} +R_i+2 R_B)}\\
%          & = \frac{1}{ \Omega C(\frac{\delta}{\Omega C_f} + Z_r \alpha l_r (1+\tan(\beta l_r)^2) +2\frac{  R_S}{2 \pi} \log (\frac{b_r}{a_r}))}\\
%          & = \frac{1}{ \Omega C(\frac{\delta}{\Omega C_f} +R_i+2 R_B)}\\
    \end{aligned}
    \label{eq12}
\end{equation}

so that the $FoM$ of Eq. (\ref{eq20}) is independent of the series capacitance. As a cross-check, it is noted that if it is assumed the $Q$ of the resonator segment is solely due to  the ferroelectric material loss and all conductor losses are neglected, Eq. (\ref{eq20}) is reduced to the $FoM$ of the pure ferroelectric as given by Eq. (\ref{eq4}).

Finally, with this circuit parametrization, the tuning range between the two  permittivity endpoints can be calculated from the reactance values, and using Eqs. (\ref{eq3}) and (\ref{eq17}) giving 
\begin{equation}
    \Delta \omega_{12} = \frac{\omega_0}{2Q_e} \frac{(X_1-X_2 )}{Z_0} 
    = \frac{\omega_0^2 Z_0}{2 Q_e} \frac{C_s^2}{C_f} \frac{\Delta \epsilon_{12}}{\epsilon_c}
    \label{eq21}
\end{equation}

Having developed the $FoM$  parametrization in terms of equivalent circuit parameters, the circuit parameters can be chosen using procedures described in the next two sub-sections.

\subsection{Determination of $Q_e$ and the Series Capacitor}

At this stage, the tuner has been defined and the framework for optimization established, but the coupling of the tuner to the cavity has not been addressed in detail. Optimization of this coupling is crucial for improving the tuner performance, as it impacts the choice of both the series capacitor $C_s$ that sets the coupling of the resonator to the transmission line, and the external quality factor $Q_e$ that sets the coupling of the full tuner system to the cavity. To address this optimization, Eq. (\ref{eq3}) can be re-expressed by introducing the a tuner specific parameter $D$

\begin{equation}
      D= \frac{X_2-X_1}{2 Z_0} 
    \label{eq22}
\end{equation}

such that  $D$ can be optimized given  the reactive power, the frequency and the number of wafers of the tuner. Then, for a given $D$, the required  tuning range sets the value of $Q_e$, which in turn, from Eq. (\ref{eq21}), sets the optimal capacitance $C_s$; namely
\begin{equation}
    C_s=\sqrt{\frac{D C_f}{\Omega_0 Z_0}  \frac{2\epsilon_c}{\Delta \epsilon_{21} }}
    \label{eq24}
\end{equation}

Again, the tuner’s resonator center frequency is required to  satisfy $\Omega_0=\omega_0$ ensuring the tuning range is centered around the cavity frequency. Since the ferroelectric capacitance  $C_f$  is  determined by the geometry, and Eq. (\ref{eq24}) sets the series capacitor $C_s$, the only adjustment left to establish the required resonance condition of the tuner  is the setting of the length of the resonator $l_r$, the short coaxial segment used to model the tuner inductance. This length $l_r$ is the combined length of the spacers and the wafers, and  using the reactance of the tuner segment from Eq. (\ref{eq7}) with $\Omega_0^2= 1/LC$ gives $l_r$:
\begin{equation}
    \tan{(\Omega_0 l_r/c}) = \frac{1}{\Omega_0 Z_r C}
    \label{eq25}
\end{equation}
 where C is the total tuner capacitance as given by Eq. (\ref{eq26}). 
\subsection{Optimization of the Figure of Merit}
To set the optimum tuner design, several details of the  resonant tuner have not yet been specified; the first of which is the radius of the inner conductor $a_r$ and the related width $w$ of the ferroelectric annulus, $w=r_a-r_i$. These two variables are related though the area of the ferroelectric wafers:
\begin{equation}
  A_{opt} = \pi N_w w (2 a_r-w)
    \label{eq27}
\end{equation}

such that a smaller width $w$ leads to a larger inner conductor radius $a_r$, implying reduced dissipation in the resonator's copper elements. The choice of actual values depends on the power, frequency and engineering constraints. It is also noted that the choice of the radii ratio $b/a$ for the transmission line has a minimal effect on the performance of the tuner and has been taken as $b/a=2$ in this work.

The second optimization is the choice of $D$ and $b_r$, but as  $D$ (which determines $C_s$) and the ratio $\theta \equiv b_r/a_r$ can not be chosen independently, a simultaneous optimization of both must be made. The existence of an optimal value for $\theta$ can be understood by inspecting the denominator of the tuner quality factor $Q$ in Eq. (\ref{eq12}). In order to maximize $Q$,  $R_i+2R_B$ must be minimized, and with  explicit expressions for $Z_r$, $\alpha$ and $l_r$ from Eq. (\ref{eq25}), the denominator can be re-expressed as a function of $\theta$ and powers of $\log(\theta)$. Indeed, the optimization of $Q$ simplifies to a minimization of the following function of $\theta$
\begin{equation}
 R_i+2R_B \propto H \frac{1+\theta^{-1}}{\log(\theta)}+\log(\theta)
    \label{eq29}
\end{equation}
where  $H=H(\Omega,C, a_r)$ is a tuner specific parameter
\begin{equation}
  H(\Omega,C, a_r) \equiv \frac{\pi c}{2 \Omega_0 ^2 C a_r \eta}
    \label{eq28}
\end{equation}

\begin{figure}[btp]
    \begin{subfigure}{\columnwidth}
        \includegraphics[width=1\columnwidth]{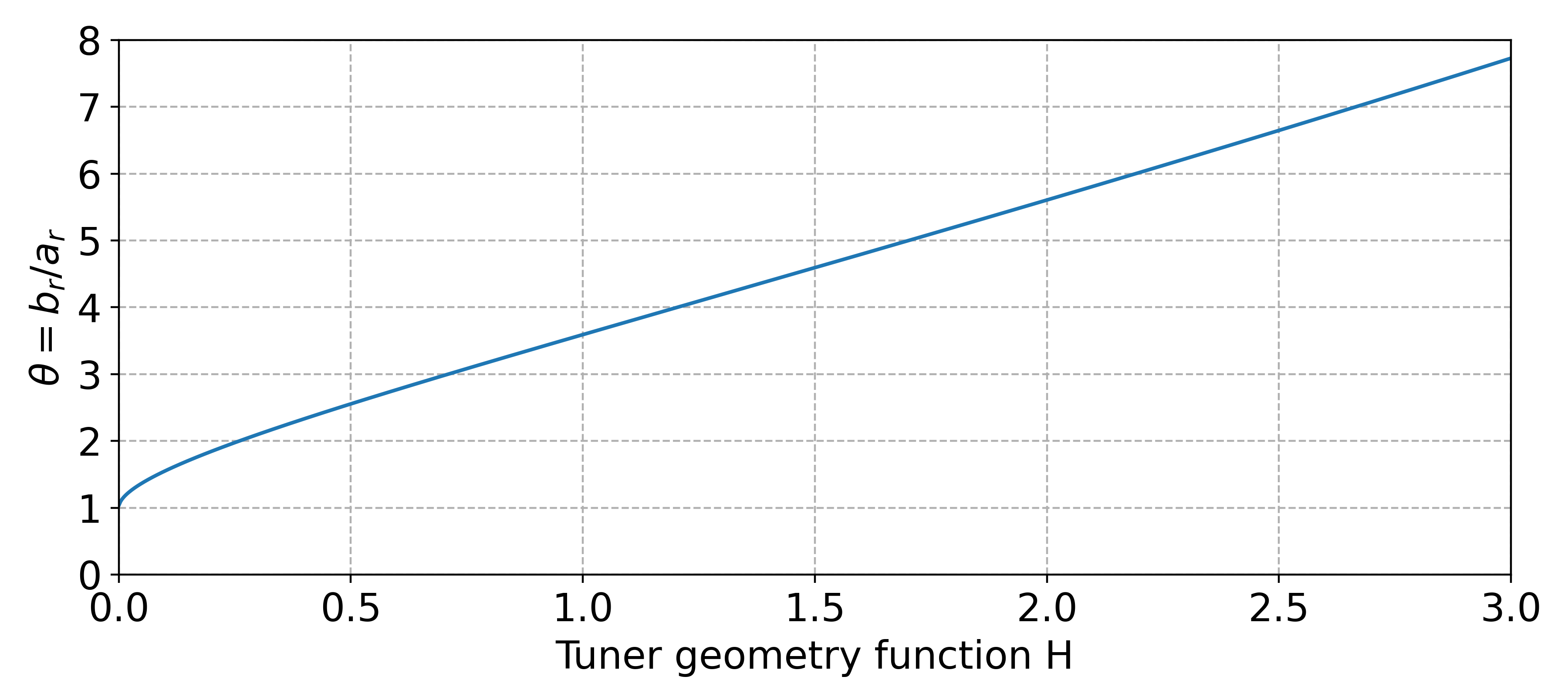	}
        \caption{The optimum  ratio $\theta=b_r/a_r$ as a function of H}
    \end{subfigure}

\bigskip

    \begin{subfigure}{\columnwidth}
        \includegraphics[width=1\columnwidth]{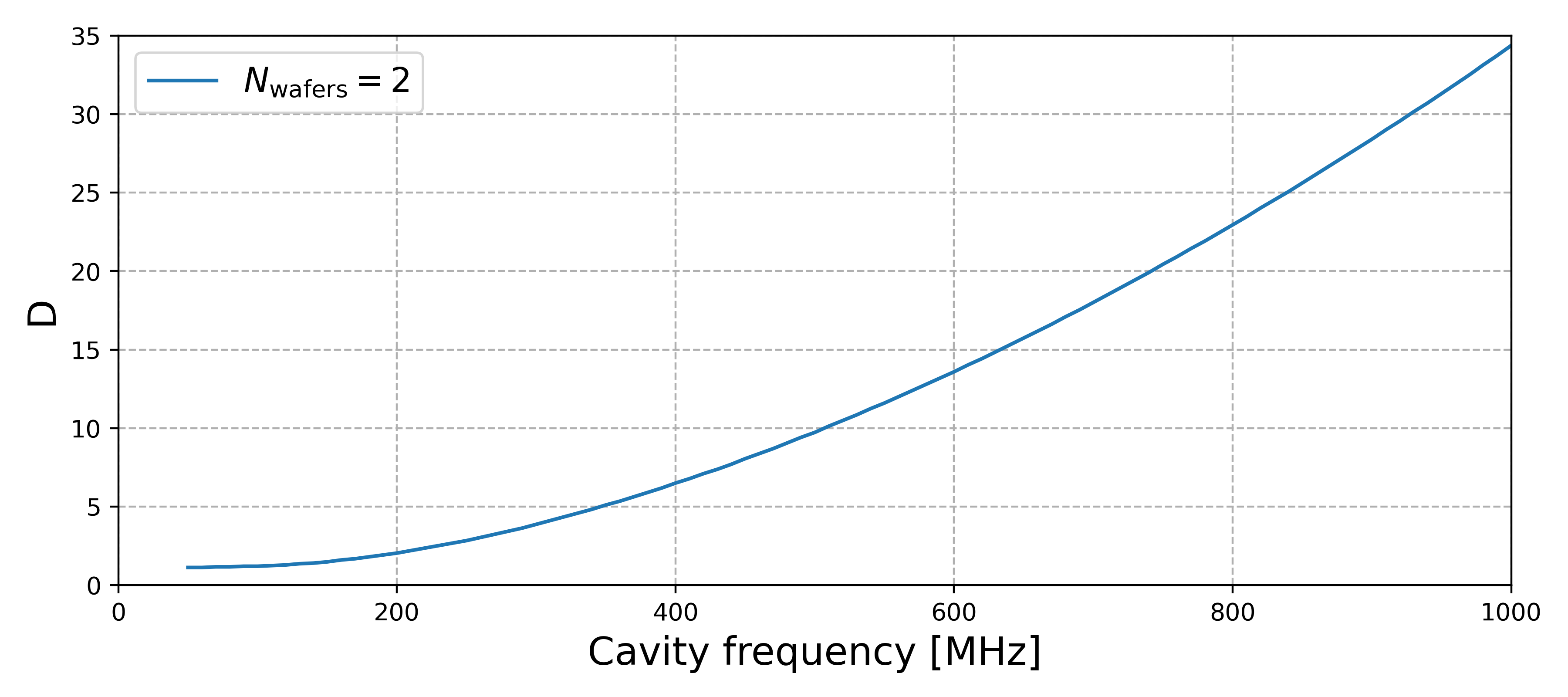}
        \caption{The optimum  $D$ at 1 MVAR and two wafers as a function of frequency}
    \end{subfigure}

\caption{\label{fig4} Optimization of the wafer aspect ratio geometry  $b_r/a_r$ and operation point D. 
} 

\end{figure}

Similarly, the existence of an optimal value of $D$ becomes clear from Eq. (\ref{eq24}): Larger $D$ increases $C_s$ and thus $C$, leading to a shorter $l_r$, implying lower inductor losses. On the other hand, if the dissipation in the quarter wavelength transmission line are considered, one can show that the $FoM$ at the cavity port is maximized when $D$ approaches $D=1$. These two competing trends lead to an optimal value of $D$, as shown in Fig. \ref{fig4}b for a 1 MVAR tuner. For a 100 kVAR tuner $D$ stays between 1 to 3.

Given that $H$ has an implicit dependence on $D$, it is clear that the optimization of $D$ and $b_r/a_r$ must be done simultaneously, and in this work, has been done numerically.

\section{Discussion and performance curves}

\begin{figure}[htp]
    \begin{subfigure}{\columnwidth}
        \includegraphics[width=1\columnwidth, height=6cm]{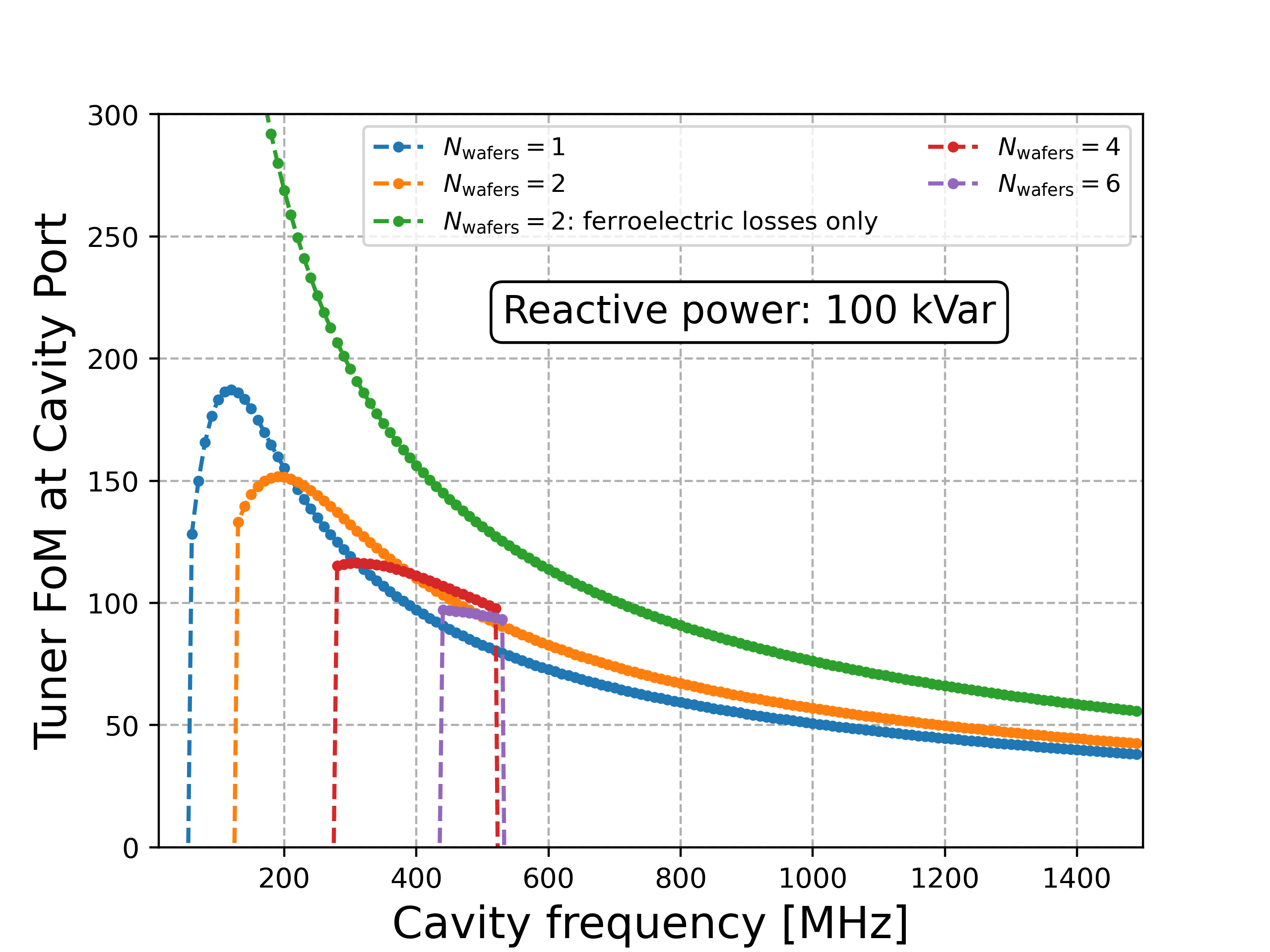}
        \caption{Tuner $FoM$ for 100 kVAR reactive power input and ferroelectric wafer 1 mm width. } 
    \end{subfigure}

\bigskip

    \begin{subfigure}{\columnwidth}
        \includegraphics[width=1\columnwidth, height=6cm]{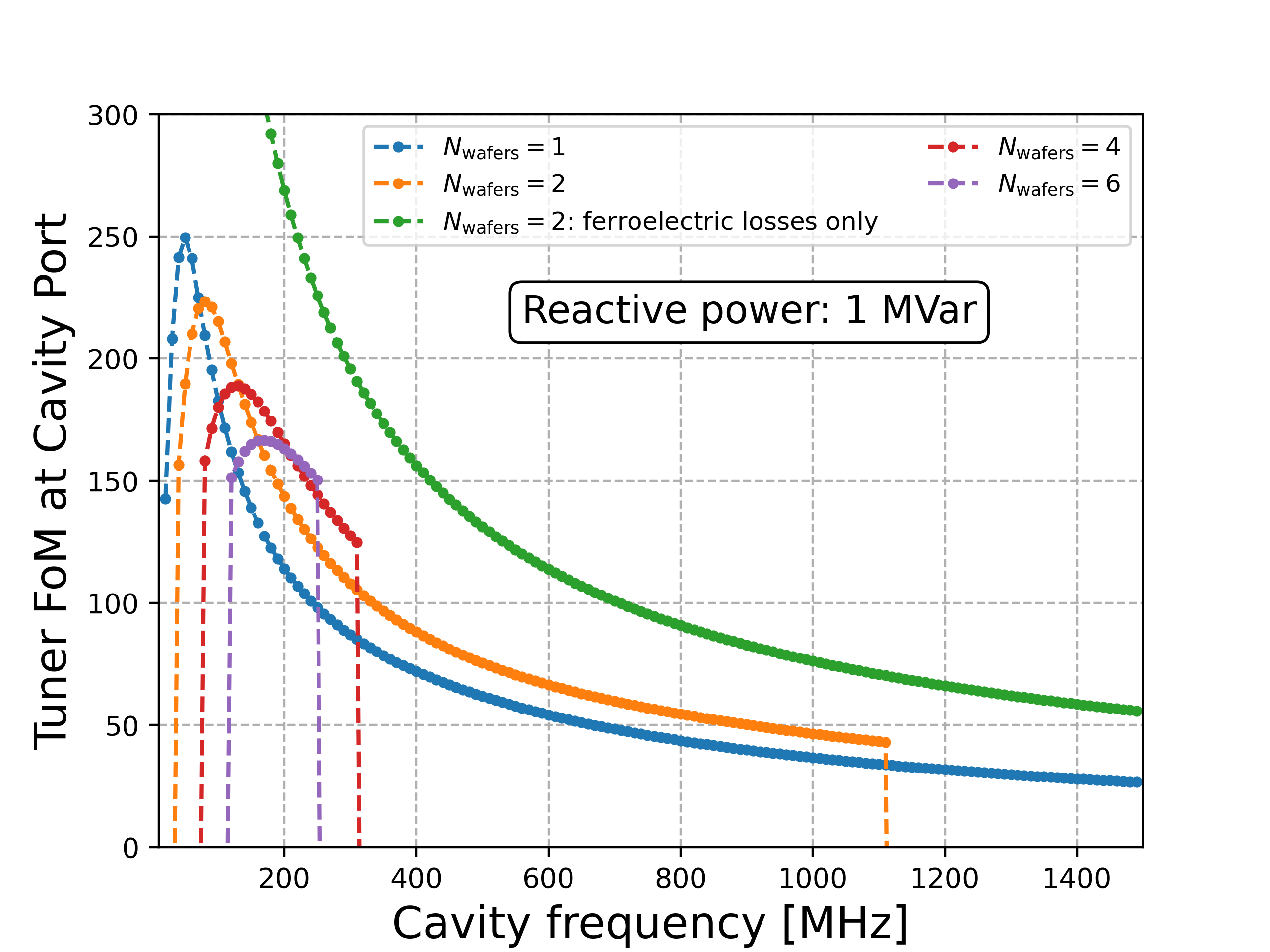}
        \caption{Tuner $FoM$ for 1 MVAR reactive power input and ferroelectric wafer 3 mm width.} 
    \end{subfigure}

\caption{\label{FoM_vs_f} The Figure of Merit of the tuner (as seen at the cavity port) as a function of frequency for two different reactive power levels. In each case the design includes optimizations described in the text. 
} 

\end{figure}

The tuner performance for an optimized tuner design can now be evaluated in terms of the tuner Figure of Merit ($FoM$) as a function of cavity frequency, and power dissipated in the tuner, and Fig. \ref{FoM_vs_f} shows the frequency dependency of the optimized $FoM$ for reactive powers of 100 kVAR and 1 MVAR, with the different curves shown corresponding to the number of stacked ferroelectric wafers. This allows the reader to quickly estimate the achievable performance of a particular tuner and match the design to the use case requirements. Given the detailed optimization discussed above, it is noted that the following design choices were made in order to generate Fig. \ref{FoM_vs_f}:  

\begin{itemize}
    \item The ferroelectric wafer geometry aspect ratio is adjusted as per Eq. (\ref{eq5}),  for each given frequency and power. The annulus width is set at 1 mm for the low power case and 3 mm for the high power case.
    \item The series capacitor is set using to Eq. (\ref{eq24})
    \item The resonator length $l_r$ is set using Eq. (\ref{eq25})
    \item The outer radius $r_f$ of the ferroelectric is equal to the radius of the spacers $a_r$
    \item The quarter wavelength transmission line to the cavity has dimensions  $a=a_r$ and $b=2 a$
\end{itemize}

It is  observed that the low frequency range for each value of $N_w$ is truncated by a requirement that the length of the resonator $l_r$ does not exceed $\lambda/8$,  so to avoid issues with the non-linear term in the resistance $R_i$ (see Eq. (\ref{eq10})). Similarly, the high frequency range is truncated at the point where  $l_r$ is less than the minimum length of the wafer spacers. In Fig. \ref{FoM_vs_f} a 0.5 cm long spacer is assumed for the two-wafer case and 1 cm long spacers for the four and six wafer designs which limits the reactive power and hence the $FoM$ at the higher frequency end of the curves.

In addition, the function $H$ has a strong dependence on frequency and tuner power (through $a_r$), and for low frequency and low power designs,  $H$ (and thus $\theta$) becomes unreasonably large. At large values of $\theta$, the frequency of the first coaxial resonator modes drop below the operating frequency which could cause unwanted mode competition. For the example shown in Fig. \ref{implementation} this occurs at $ \approx \theta = 7$. Consequently, the results shown in Fig. \ref{FoM_vs_f}(a) include a limit of $H<3$, and the $FoM$ at low frequencies is not fully optimized, as is evident from a comparison with Fig. \ref{FoM_vs_f}(b).

As these figures represent tuner performance over a considerable frequency range, ferroelectric wafer configurations and widely different reactive power levels, some immediate conclusions may be drawn.  For tuner designs with in series stacks of ferroelectric wafer geometries,  designs with more than  2 wafers does not lead to an advantage in $FoM$ at low reactive power levels; with a 4-wafer design, there may be a small performance advantage at low frequencies, but at the price of added mechanical complication.

The $FoM$ values, following optimization procedures, are relatively high, being of the order of 100 at 100 kVAR reactive power. At the higher reactive power of 1 MVAR, the $FoM$ starts higher at low frequencies, even exceeding 200, but drops fast with increasing frequency.

\section{Electromagnetic design procedure}\label{section4}
\begin{figure}[!htb]
    \centering
    \includegraphics[width=0.95\columnwidth]{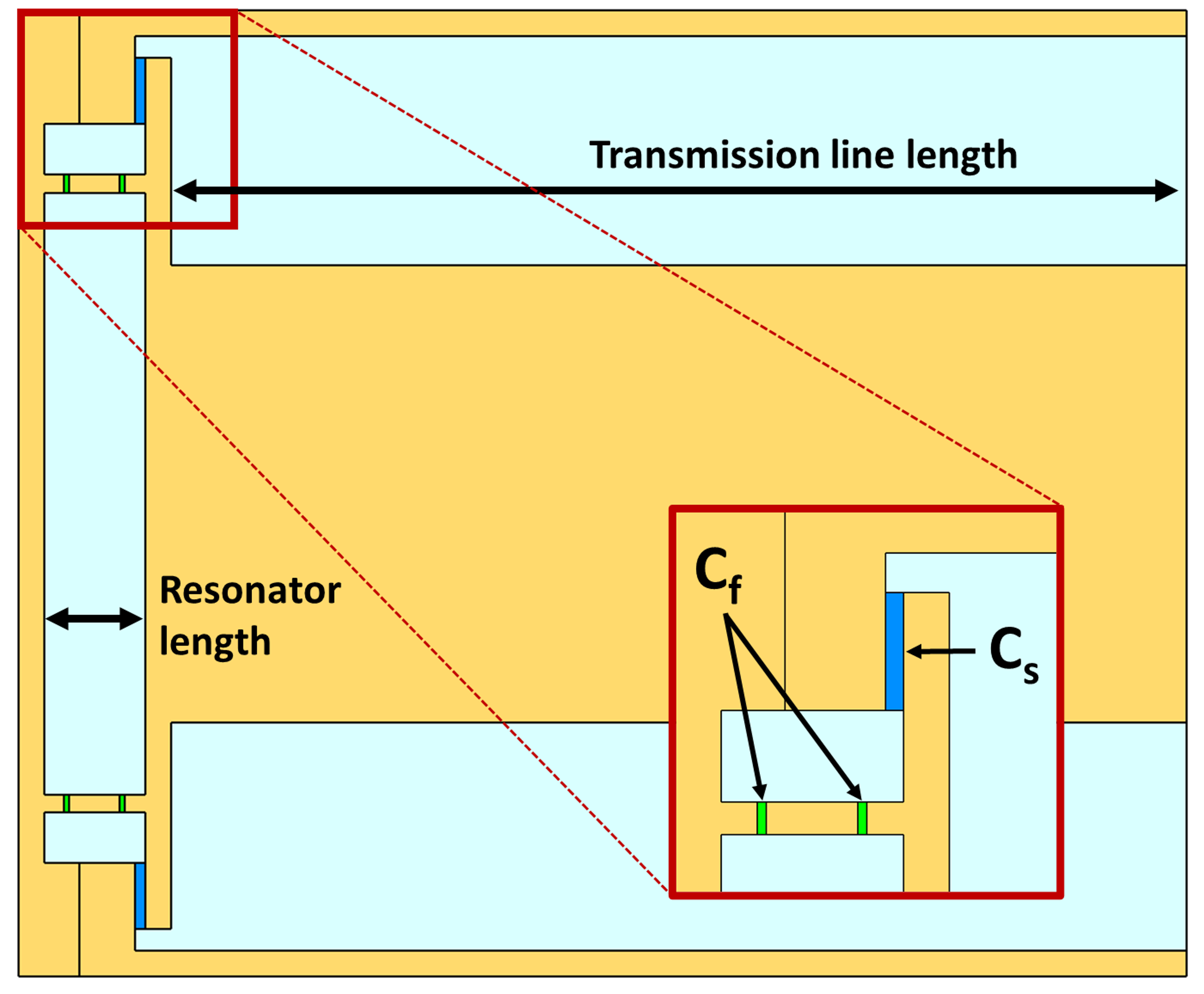}
    \caption{\label{implementation}CST model of tuner design showing the implementation of $C_f$ and $C_s$. Light blue represents vacuum space.}
\end{figure}

In order to cross-check the analytical formulas from the previous sections an assessment of tuner design performance based on  a detailed modelling and electromagnetic simulation has been performed, using the  CST Studio Suite \cite{CST} simulation package. The tuner design example is as summarized by the analytically calculated input parameters given in Table \ref{tab:table2}, and represents a tuner design comprising two ferroelectric wafers of an annulus shape, capable of sustaining 1.9 MVAR reactive continuous tuning power. An outline of the design is shown in Fig. \ref{implementation}, with the two ferroelectric wafers that comprise the capacitance $C_f$,  and a  (non-ferroelectric) sapphire annulus as the capacitance $C_s$ shown in the insert of Fig. \ref{implementation}. Implementing $C_s$ in this way also provides an clean way to separate the resonator vacuum and the cavity vacuum. The top part of the resonator shows a break indicating that it is removable and would be sealed with a vacuum flange. It is noted that this model is not a full mechanical design of a tuner and only a representative model for electromagnetic simulation purposes.

\begin{table}[b]
\begin{ruledtabular}
\begin{tabular}{ccccc}
Parameter  &Analytical& Units\\
\hline
$f_0$ & 400.7948 & MHz \\
$U$ & 136 & J \\
$\Delta f$ & 1.1 & kHz \\
$A_{opt}/g$ & 1153.7 & mm \\
w & 3 & mm \\
$Z_0$ & 41.6& $\Omega$ \\
a & 45 & mm\\
b & 90 & mm \\
$a_r$ & 62.71 & mm \\
$b_r$ & 72.77 & mm\\
$Q_e$ & 3.25 & $\times 10^6$  \\
\end{tabular}
\end{ruledtabular}
\caption{Analytical design parameters used for comparison. \label{tab:table2}}

\end{table}

\subsection{Stand Alone Tuner}
\begin{figure}[!htb]
    \centering
    \includegraphics[width=0.95\columnwidth]{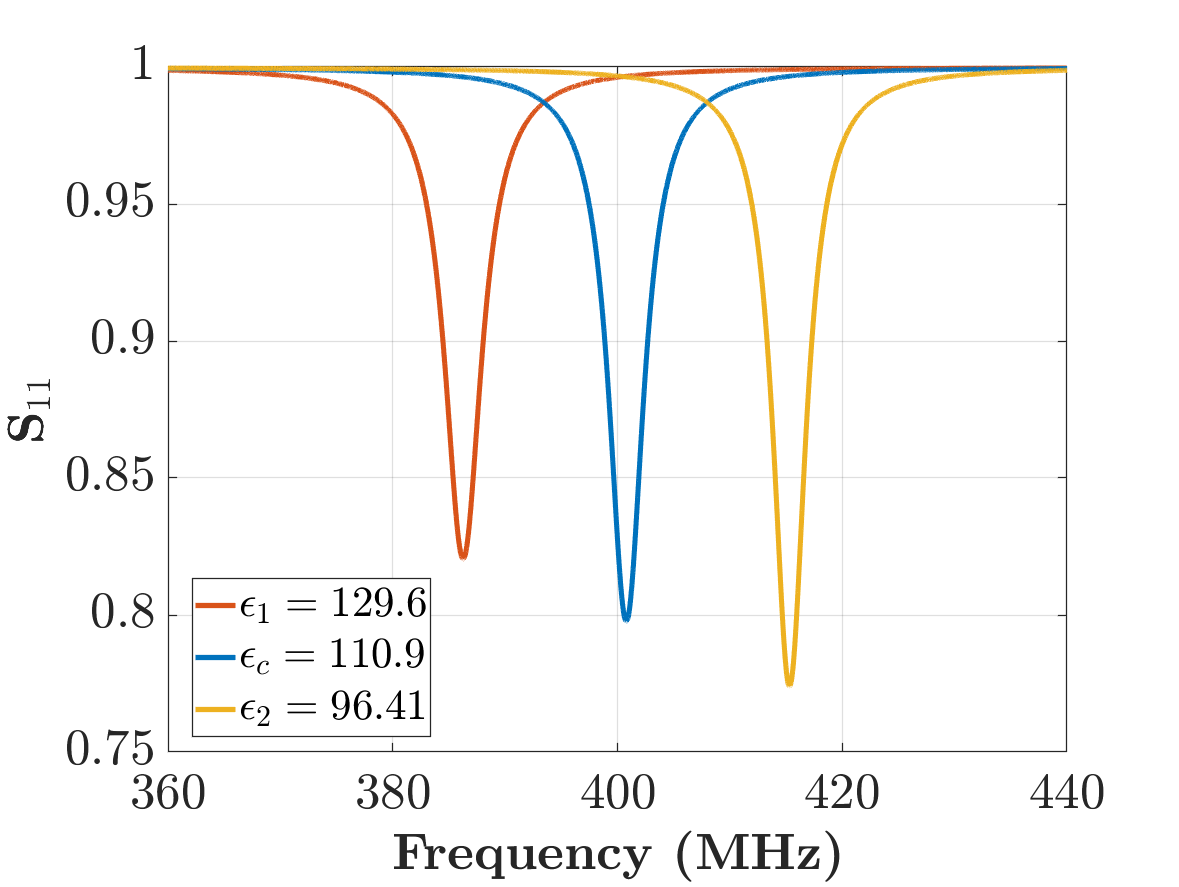}
    \caption{\label{S11vsfreq} $S_{11}$ vs. frequency for the two end states ($\epsilon_1$, $\epsilon_2$) , and a central value of $\epsilon_c$.}
\end{figure}

To implement the detailed modelling of this high power tuner design for a 400 MHz cavity, the value of the $A_{opt}/g$ was set to the same value as the analytical model, and is defined by the minimum area that can sustain the power dissipated in the material. In addition, the width of the sapphire capacitor $C_s$ was initially set using an electrostatic solver to estimate the capacitance. With these two values set, the model was then generated and simulated, with the tuner geometry as shown in Fig. \ref{implementation}.  The permittivity of the ferroelectric material was varied from $\epsilon_1$ to $\epsilon_2$, with the $S_{11}$ response observed for each state. The tuner resonant frequency for each state was evaluated, and then the resonator length was set such that the the average of the two end state frequencies was at the cavity frequency, and so that the end states were equally spaced around $f_0$. A central value of the permittivity that gives this resonance at $f_0$ can be defined, as shown in Fig. \ref{S11vsfreq}, along with the two end state resonances.  

\subsection{Attaching the Tuner to a Cavity}
\begin{figure}[b]
    \centering
    \includegraphics[width=1\columnwidth]{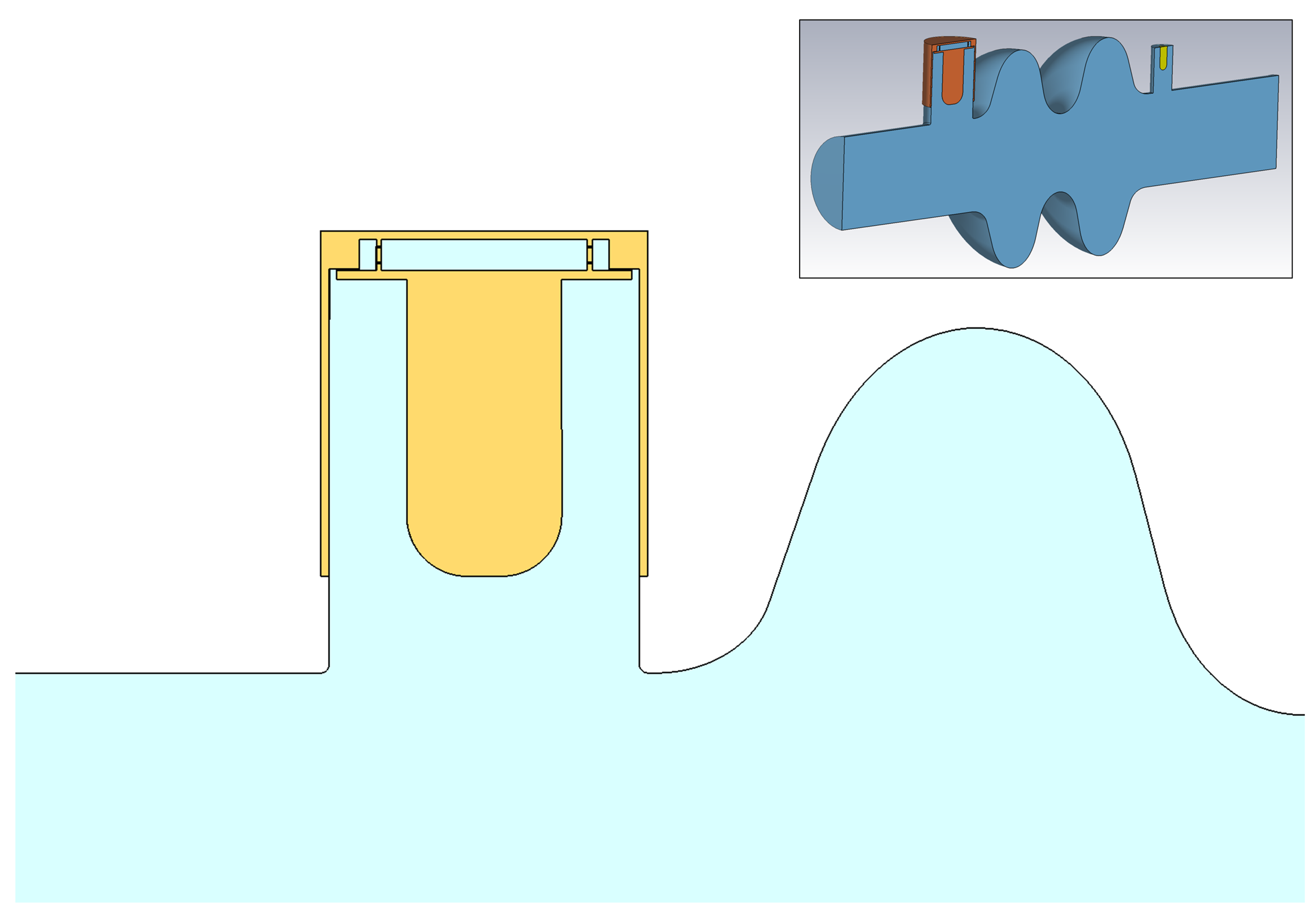}
    \caption{\label{tuner+cav} Detail of tuner attached to two cell 400 MHz cavity.}
\end{figure}

After the dimensions of the tuner had been approximately set, it was attached to a 2-cell, 400 MHz cavity as shown in Fig. \ref{tuner+cav}. The cavity had three ports, one for the FRT with $Q_e$ set to the design value from Table \ref{tab:table2}, a fundamental power coupler (FPC) port with $Q_e = 5\times 10^7$  and a probe port with $Q_e = 5\times 10^9$. The stand alone cavity with the three ports, including the FRT port was simulated separately in order to set the values of $Q_{e}$, where each port is terminated with its characteristic impedance. The resonant frequency of the three port cavity was also set to $f_0$, assumed to be the resonant frequency with no tuner attached.

As per the analytic model, the transmission line length connecting the tuner to the cavity should be exactly $\lambda/4$ in length, though uncertainty in the region where the coaxial line transitions into the cavity means this is difficult to estimate. In order to properly set the transmission line length, the tuning range response as a function of  line length was simulated, and the minima of the tuning range used to identify  the correct transmission line length for that particular tuner configuration. Further, it was found that with the input parameters from the analytical model, the tuning range minima was only 0.97 kHz instead of the 1.1 kHz required.  To increase the tuning range without affecting the other tuner parameters, $Q_e$ was adjusted from $3.25\times 10^6$ to $2.95\times 10^6$,  bringing the tuning range to 1.1 kHz. Finally, to have  a clear comparison quantity for the CST simulation, $Q_{FRT}$ was calculated using the known $Q_e$ of the FPC and probe ports and the measured $Q_L$ using: 

\begin{equation}
   Q_{FRT} = \frac{1}{\frac{1}{Q_L} - \frac{1}{Q_0} -\frac{1}{Q_{FPC}} - \frac{1}{Q_{probe}} }
\end{equation}

The $FoM$ was then evaluated using the average $Q_{FRT}$ of the two end states and the tuning range, $\Delta f$:

 \begin{equation}
   FoM = \frac{4\pi\Delta f \overline{Q_{FRT}}}{\omega_0}
\end{equation}

and the 12\% reduction in $Q_e$ was expected to cause a similar reduction in the measured $Q_{FRT}$ and hence the FOM.
\begin{figure}[htb]
    \centering
    \includegraphics[width=0.95\columnwidth]{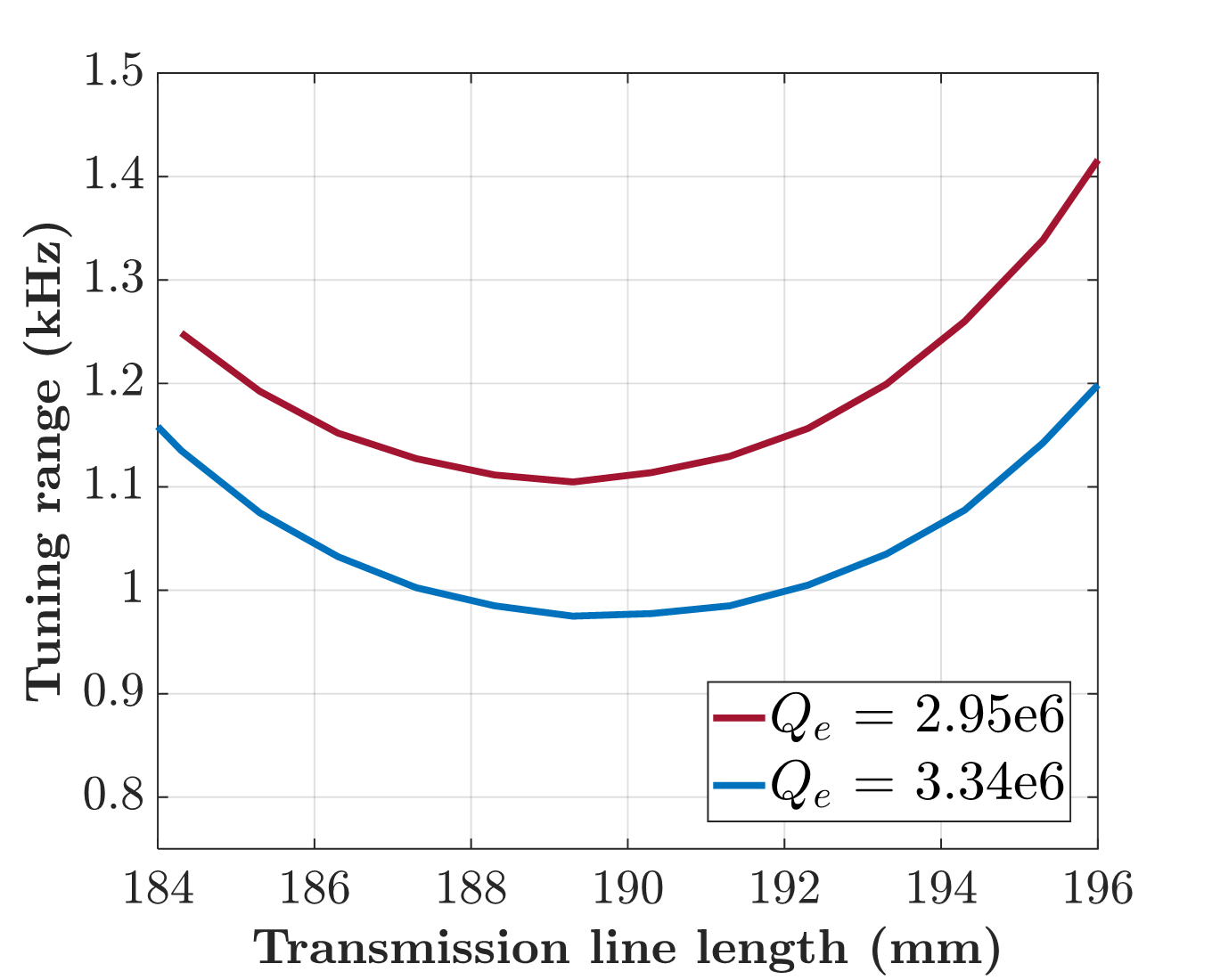}
    \caption{\label{TRvsTL} Tuning range vs. transmission line length for two values of $Q_e$.}
\end{figure} 

\subsection{Final Tuning}
Final fine adjustment of parameters to center the  end states around the cavity frequency $f_0$ while maintaining the required tuning range was performed using an the in-built CST optimization framework, with the following optimization objectives:
\begin{itemize}
    \item[A:] $\Delta f = | f_1 - f_2 | = 1.1 \hspace{1mm}\text{kHz}$
    \item[B:] $\delta_f = |f_{0}-(f_1+f_2)/2| < 0.01 \hspace{1mm}\text{kHz}$
\end{itemize}

where $f_1$ and $f_2$ are the frequencies of the two end states. Clearly, Objective A is the tuning range requirement, while Objective B enforces the tuning range to be centered on the cavity frequency with no tuner attached, $f_0$.

\begin{table}[bt]
\begin{ruledtabular}
\begin{tabular}{ccccc}
Parameter  &Analytical& CST & Units\\
\hline
$\Delta f$ & 1.1 & 1.1 & kHz\\
$C_{f}$ & 551.9 & 563.3 & pF \\
$C_s$ & 578.3 & 584 & pF\\
$l_r$ & 18.39 & 18.05 & mm \\
$\overline{Q_{FRT}}$ & $3.18$ &  $2.38$ & $\times 10^7$\\
$FoM$ & 81.6 & 65.3 & - \\
df1 & 550.3 & 554.8 & Hz\\
df2 & 549.8 & 550 & Hz\\
$Q_e$ & $3.25$ & $2.95$ & $\times 10^6$ \\
\end{tabular}
\end{ruledtabular}
\caption{Comparison of analytical and optimized CST values for a 400 MHz tuner design. \label{tab:table3}}
\end{table}

\begin{figure}[!htb]
    \centering
    \includegraphics[width=1\columnwidth]{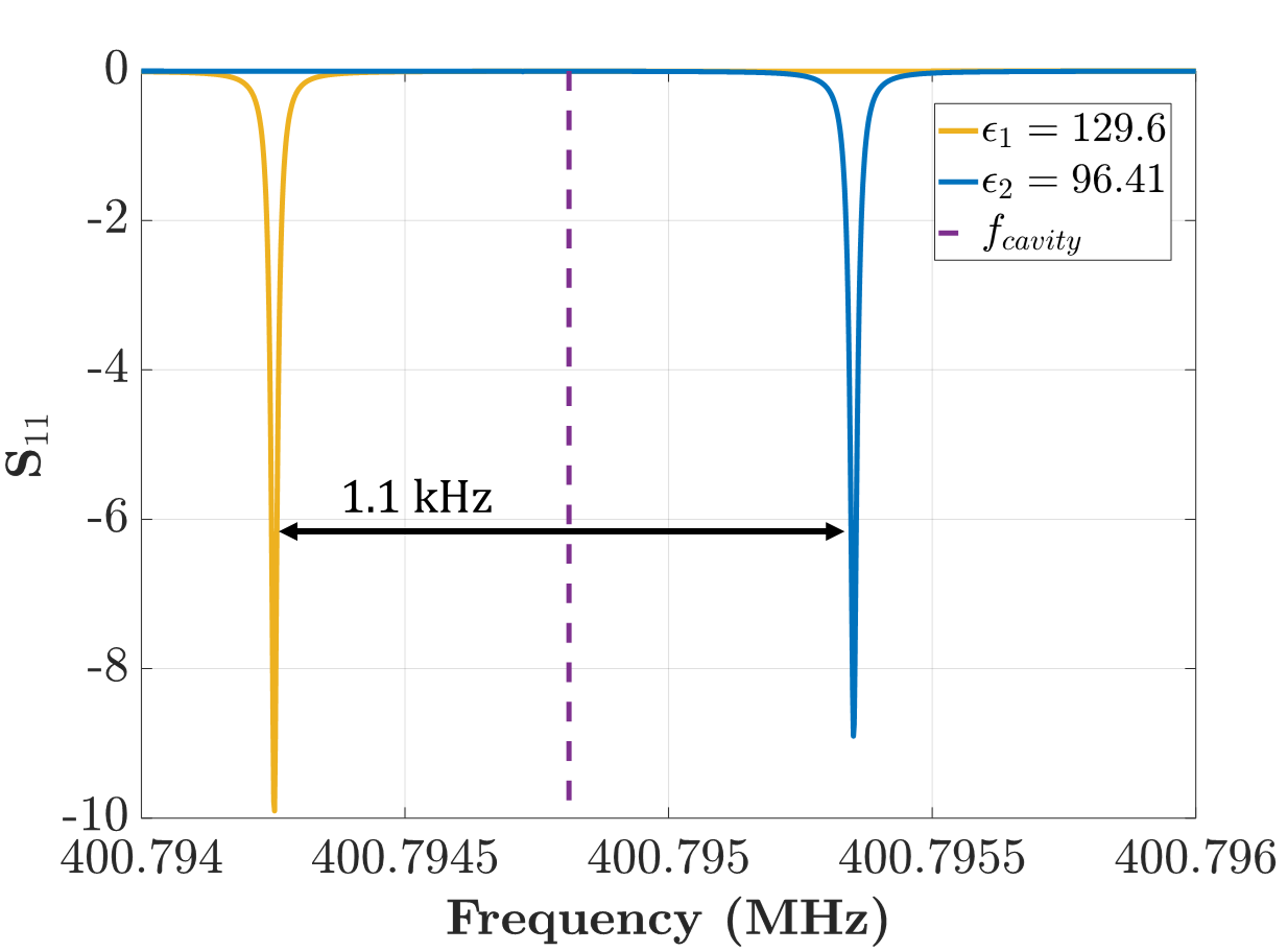}
    \caption{\label{final_sparams} Final S-parameters observed through the FPC port with $Q_e = 5\times 10^7$.}
\end{figure}

The inputs of the optimization were the geometry of the sapphire annulus making up $C_s$ and the tuner resonator length, $l_r$, and the comparison of the analytical and CST simulation results are shown in Table \ref{tab:table3}. Excellent agreement is seen between the analytical capacitance values and those estimated using an electrostatic solver. The length of the resonator also shows good agreement, indicating that the analytical model predicts the frequency of the tuner resonator well. The final S-parameters as observed through the FPC port are shown in Fig. \ref{final_sparams}. It is noted that the analytical model predicts a larger value of $\Delta X_{12}$ than is obtained in CST for the particular tuner configuration. This implies to a need to reduce $Q_{e}$ in the CST model, in order to increase the tuning range, and correspondingly, leads to a reduction in the CST values of both $Q_{FRT}$ and the $FoM$. If $Q_e$ is left unchanged, $Q_{FRT}$ can be increased to $3.1\times 10^7$, which is more in line with the analytic model,  but the tuning range is reduced to $0.93$ kHz. Finally, including all the loss mechanisms that are not included in the analytical model such as extra lengths of copper and the losses in the sapphire used to make $C_s$, further reduces $Q_{FRT}$ by $\approx$ 4\%. 

The same methodology was also applied to an 800 MHz tuner design at a power level of 458 kVAR, with the analytical values presented in Table \ref{tab:800MHz1} and comparison in Table \ref{tab:800MHz}. This tuner design was based on a potential use-case for a conceptual FCC-ee booster scenario whereby cavities must be detuned by 1.5 kHz every 3.8s during injection \cite{fcc2019fcc}. For this case $Q_e$ was not adjusted to compensate for the decreased tuning range compared to the analytical estimate, resulting in a $\approx$ 15 \% reduction in tuning range. The $FoM$ is also much closer to the analytical value with the $\sim 13\%$ difference being attributed to extra losses that are not accounted for in the analytical model. The agreement between the analytical and numeric calculations is excellent just as in the 400 MHz case.

\begin{table}[htb]
\begin{ruledtabular}
\begin{tabular}{ccccc}
Parameter  &Analytical & Units\\
\hline
$f_0$  & 801.59& MHz \\
$U$ & 24.3 & J \\
$\Delta f$ & 1.5 & kHz \\
$A_{opt}/g$ & 483.2 & mm \\
w & 2 & mm \\
$Z_0$ & 41.6 & $\Omega$ \\
a & 32.5  &  mm\\
b & 65 & mm \\
$a_r$ & 39.5 & mm \\
$b_r$ & 45.5 & mm\\
$Q_e$ & 4.9 & $\times 10^6$ \\
\end{tabular}
\end{ruledtabular}
\caption{Analytical design parameters (800 MHz). \label{tab:800MHz1}}
\end{table}

\begin{table}[t]
\begin{ruledtabular}
\begin{tabular}{ccccc}
Parameter  &Analytical& CST & Units\\
\hline
$\Delta f$ & 1.5 & 1.3 & kHz \\
$ C_{f}$ & 230.9 & 239.5 & pF\\
$ C_{s}$ & 268 & 280.6 & pF\\
$l_r$ & 10.8 & 8.59 & mm \\
$\overline{Q_{FRT}}$ & $3.28$ &  $3.1$ & $\times 10^7$\\
$FoM$ & 56.6 & 49.3& - \\
df1 & 755 & 737 & Hz\\
df2 & 745 & 548& Hz\\
\end{tabular}
\end{ruledtabular}
\caption{Analytical vs. CST design (800 MHz). \label{tab:800MHz}}
\end{table}

\section{SUMMARY}\label{section5}

This paper presents a methodology for designing and optimizing a coaxial type resonant ferroelectric fast reactive tuner operating at high average reactive power levels over a broad range of operating frequencies. The tuner as represented by the schematic of Fig. \ref{Circuit}, comprises a ferroelectric capacitor, which has $N_W$ wafers in series, separated by spacers which provide a bias voltage and cooling,. A particular ferroelectric material is used in the design, with material properties taken from the ferroelectric supplier \cite{kanareykin2005low}. 

Given a cavity with a known stored energy, operating frequency and desired tuning range, the reactive power for the tuner is determined. The ferroelectric material properties together with the frequency and reactive power determine the minimum area of the ferroelectric wafers that can sustain this power. The choice of ferroelectric ring geometry and sizing  then leads to a choice of the radius of the inner conductor of the tuner resonator segment, and the optimization of the $FoM$ leads to specific choices for the coupling capacitor $C_s$ and the radius of the resonator's outer conductor. Depending on the frequency and reactive power requirements, Fig. \ref{fig4} then gives the preferred number of ferroelectric capacitors in series $N_w$. 

To illustrate the application of this design methodology to a particular case, two examples are considered, with analytic modeling of optimized tuner design compared to full electromagnetic simulation of 3-dimensional geometry. The first is a case of a very demanding tuner for a two cell 400 MHz cavity with the significant reactive power level of $\Delta P_{\text{reactive}}=1.9$ MVAR. The second is a higher frequency (800 MHz) tuner with a reactive power level of 458 kVAR. The agreement is excellent for both examples, proving that the analytical model provided gives an accurate set of initial parameters before starting on the detailed 3D electromagnetic design.

\begin{acknowledgments}
This work is in part supported through the  Innovate for Sustainable Accelerating Systems  (iSAS) programme funded through the European Commission’s Horizon Europe Research and Innovation programme under Grant Agreement n°101131435. 

I.B-Z. acknowledges support under the CERN Visiting Scientist program.

\end{acknowledgments}

\bibliography{bib}

\end{document}